\documentclass[slac_one]{revtex4}
\usepackage{graphicx}
\usepackage{fancyhdr}
\usepackage{amsfonts}
\begin{document}

\title{The structure of $\mathcal{N}=2$ multi-instanton and string-loop corrections in toroidal orbifold models}

\author{P. G. C\'amara}
\affiliation{Centre de Physique Th\'eorique, Ecole Polytechnique,
CNRS, F-91128 Palaiseau, France}

\begin{abstract}
We summarize the structure of $\mathcal{N}=2$ multi-instanton and
string loop corrections to the K\"ahler potential and the gauge
kinetic function in type I toroidal orbifold models. For that aim,
we exploit perturbative calculations of the physical gauge
couplings on the heterotic dual, whenever such a perturbative
description is available.
\end{abstract}

\maketitle

\thispagestyle{fancy}

\section{INTRODUCTION}

The low energy limit of String Theory compactified to four
dimensions is often described in terms of an effective
supergravity which generically develops classical flat directions.
Understanding the leading corrections from string loops and
non-perturbative effects is therefore crucial for addressing
questions such as moduli stabilization, supersymmetry breaking or
cosmological evolution in these models. Although a
non-perturbative description of String Theory is still far from
being at our disposal, the rich network of dualities allows,
already at this stage, to make some explicit computations in
simple models. Precisely, type I/heterotic S-duality \cite{sdu}
maps stringy instanton effects due to euclidean 1-branes
(E1-branes), into string loop effects in a dual heterotic model.
Whereas in ten dimensions S-duality is a strong-weak coupling
duality, in four dimensions it is manifested as an
electric-magnetic duality, making the mapping between both
effective theories straightforward.

In this contribution we mainly summarize some of the results of
\cite{cd}, where the structure of multi-instanton and string loop
corrections in a simple class of models were obtained. Concretely,
we consider type I string theory compactified on toroidal
orbifolds, and use type I/heterotic S-duality to extract the
corrections to the K\"ahler potential and the gauge kinetic
function, coming from string loop and multi-instanton effects
associated with $\mathcal{N}=2$ subsectors of the theory.
Additional related material and a more detailed exposition can be
found in \cite{cd,cd2}.

A simple way to extract corrections to the K\"ahler potential,
$K$, and the gauge kinetic function, $f_a$, of the effective
theory is to look at the one-loop physical gauge couplings. In a
heterotic compactification, these read,
\begin{equation}
4\pi^2g_a^{-2}(\mu^2)|_{\rm
1-loop}=\frac{k_a}{\ell}+\frac{b_a}{4}\textrm{log}\frac{M_{\rm
s}^2}{\mu^2}+\frac{\Delta_a(M,\bar M)}{4}\ ,\label{thresgeneral}
\end{equation}
with $\ell$ the linear multiplet associated with the dilaton,
$M_s$ the string scale, $M$ the moduli of the compactification and
$k_a$ the normalization of the gauge group generators, determined
by the level of the corresponding Kac-Moody algebra. The
$\beta$-function coefficient, $b_a$, is given in terms of the
quadratic Casimir invariants of the gauge group, $b_a=\sum_r
n_rT_a(r)-3T_a({\rm adj}_a)$, with $n_r$ the number of matter
multiplets in the representation $r$.

On the other hand, in field theory the physical gauge couplings
are given as \cite{kl1,kl},
\begin{eqnarray}
16\pi^2g_a^{-2}(\mu^2)|_{\rm 1-loop}&=&4\textrm{Re }f_a(M)|_{\rm
1-loop}+b_a\textrm{log}\frac{M_{\rm Pl}^2\mu^{-2}}{S+\bar S}+
c_a\hat K(M,\bar M)-2\sum_rT_a(r)\textrm{log det }Z_r(M,\bar M)\ ,
\label{field}
\end{eqnarray}
where det $Z_r$ is the determinant of the tree-level K\"ahler
metric associated with the matter multiplets in the representation
$r$, $\hat K(M,\bar M)$ the tree-level K\"ahler potential for the
moduli $M$ and $c_a=\sum_rn_rT_a(r)-T({\rm adj}_a)$.

In order to compare (\ref{thresgeneral}) and (\ref{field}), we
recall the relation between the usual complex axion-dilaton $S$
and the linear multiplet $\ell$,
\begin{equation}
\frac{1}{\ell}=\textrm{Re }S-\frac{1}{4}\Delta_{\textrm{univ.}}
\end{equation}
with $\Delta_{\textrm{univ.}}$ a gauge group independent
(``universal'') function which, for convenience, we split in to
its harmonic and non-harmonic parts,
\begin{equation}
\Delta_{\textrm{univ.}}(M,\bar M)=V_{(1)}(M,\bar M)+H(M)+H^*(\bar
M)\ .
\end{equation}
The K\"ahler potential and the gauge-kinetic function of the
effective theory then are given to one-loop by
\cite{derending,kl,stiebergerii}
\begin{eqnarray}
K|_{\rm 1-loop}&=&-\textrm{log}\left(S+\bar
S-\frac12V_{(1)}(M,\bar M)\right)+\hat K(M,\bar M)\ ,\label{k1}\\
\textrm{Re }f_a|_{\rm 1-loop}&=&k_a \textrm{Re }S +
\frac{1}{4}\left(\Delta_a(M,\bar M) - V_{(1)}(M,\bar M)- c_a\hat
K(M,\bar M) -2\sum_rT_a(r)\textrm{log det }Z_r(M,\bar M)\right)\
.\label{k2}
\end{eqnarray}

\section{THE MODELS}

We focus on type I toroidal orbifold compactifications which admit
a weakly coupled perturbative description in terms of heterotic
string theory. More precisely, we consider models where the
orbifold action, $\mathbb{G}$, contains some subgroup,
$\mathbb{G}_i$, which leaves unrotated some complex plane, and
therefore there is a $\mathcal{N}=2$ sector of the theory
associated to each of such subgroups. This includes
$\mathbb{Z}_2$, $\mathbb{Z}_2\times\mathbb{Z}_2$, $\mathbb{Z}_6$,
$\mathbb{Z}_6'$, $\mathbb{Z}_8$ or $\mathbb{Z}_{12}$ orbifold
models.

The gauge threshold corrections can be computed in the heterotic
side along the lines of \cite{kaplu,dkl,kk2,kk3} and take the
expression,
\begin{equation}
\Delta_{a}=-b_a\textrm{log}\frac{M_{\rm
s}^2}{\mu^2}-\frac12\sum_i\int_{\mathcal{F}}\frac{d^2\tau}{\tau_2}\hat
Z_i(T_i,U_i)\hat{\mathcal{A}}^a_{f,i}\ ,\label{integral}
\end{equation}
where the sum runs over the disjoint union of $\mathcal{N}=2$
subsectors, each leaving invariant a single complex plane with
K\"ahler and complex structure moduli $T_i$ and $U_i$, and the
gauge group is given by a product $G=\prod_a G_a$. The explicit
expression for the toroidal lattice sums, $\hat{Z}_i$, can be
found in \cite{cd}. The index $\hat{\mathcal{A}}^a_{f,i}$ is an
almost holomorphic modular function which is constrained by
modular invariance and $\mathcal{N}=2$ supersymmetry to have the
structure,
\begin{equation}
\hat{\mathcal{A}}^a_{f,i}(\tau)=2b_i^a+\frac{\gamma_i}{20\eta^{24}}\left[D_{10}E_{10}-528\eta^{24}\right]\
,\label{index}
\end{equation}
with $b_i^a$ the $\beta$-function coefficient of the
$\mathcal{N}=2$ gauge theory associated to a would-be
$T^6/\mathbb{G}^i$ orbifold and $\gamma_i$ a gauge group
independent coefficient. Definitions for the Eisenstein series,
$E_k$, and the modular covariant derivative, $D_k$, can be found
e.g. in the appendix of \cite{cd2}.

For the sake of clarity, let us briefly consider the simplest case
of a single $\mathcal{N}=2$ sector. This corresponds to the well
known case of type I string theory compactified on $T^2\times K3$.
In order to cancel anomalies, 24 instantons are required. In the
$T^4/\mathbb{Z}_2$ orbifold limit of the $K3$ \cite{gp1,gp2} these
are manifest as 8 D5-branes and 16 $U(1)$ bundles hidden at the 16
orbifold singularities. The D5-branes together with the 16
D9-branes also present, lead to a maximal gauge group $U(16)\times
U(16)$.

In the Coulomb branch of this setup, where a half D5-brane is
located at each of the 16 fixed points, only the $U(16)$ gauge
group from the D9-branes remains massless because of the
Green-Schwarz mechanism. The spectrum is given by four
hypermultiplets, containing the moduli of the $K3$, three vector
multiplets containing the axion-dilaton and the moduli of the
$T^2$, a $\mathbf{120}+\overline{\mathbf{120}}$, and sixteen
$\mathbf{16}$ from the D5-D9 modes. In this particular point of
the moduli space the weakly coupled SO(32) heterotic dual
therefore does not contain $NS5$-branes, and can be worked out
explicitly \cite{polcho,luis}. This is given by a $T^2\times
T^4/\mathbb{Z}_2$ orbifold with shift vector $V=\frac14
(1,\ldots,1,-3)$, corresponding to $(b_{U(16)},\gamma)=(12,-1)$ in
eq.(\ref{index}).

More generically, the condition $\gamma=-1$ is related to points
in the moduli space where a dual description can be given in terms
of perturbative $E_8\times E_8$ heterotic string compactified on a
$T^6$, with no Wilson lines, and instanton numbers $(12+n,12-n)$,
with $n=2+\frac{b_a}{3}$ \cite{stiebergeriii}. In the above
Coulomb branch, we hence obtain instanton numbers $(18,6)$, and we
are sitting at the point where the moduli spaces of perturbative
SO(32) and $E_8\times E_8$ heterotic strings intersect.

\section{PERTURBATIVE AND NON-PERTURBATIVE CORRECTIONS}

The integral (\ref{integral}) can be evaluated using the methods
of \cite{dkl}, which consists of dividing $\hat Z_1$ into orbits
under the modular group and evaluating the resulting integral for
each orbit representative in a suitable unfolded region of the
upper complex half-plane. Putting all pieces together and making
use of eqs.(\ref{k1}) and (\ref{k2}), we obtain the main result of
this note \cite{cd},
\begin{eqnarray}
K&=&-\textrm{log}(S+\bar S)-\sum_{i}\bigg\{\textrm{log}[(T_i+\bar
T_i)(U_i+\bar U_i)]+\frac{1}{2}\frac{V^i_{1-loop}+V^i_{E1}}{S+\bar
S}\bigg\}+\ldots\ , \\
V^i_{1-loop}&=&\frac{4\pi \gamma_i}{3}\frac{E(iU_i,2)}{T_i+\bar T_i}\ , \label{kloop}\\
V^i_{E1}&=&\frac{\gamma_i}{\pi}\sum_{k>j\geq 0,\
p>0}\frac{e^{-2\pi kp
T_i}}{(kp)^2}\left[\frac{\hat{\mathcal{A}}_K(\mathcal{U}_i)}{T_i+\bar
T_i}-\frac{2i
kp}{\mathcal{U}_i-\bar{\mathcal{U}_i}}\frac{E_{10}(\mathcal{U}_i)}{\eta^{24}(\mathcal{U}_i)}\right]\
+ \ \textrm{c.c.}\ , \\
f_a&=&S+\sum_i\bigg\{\frac{\pi
(b_i^{a}+6\gamma_i)T_i}{12}-b^a_i\textrm{log
}\eta(iU_i)-\frac12\sum_{k>j\geq 0,\ p>0}\frac{e^{-2\pi
kpT_i}}{kp}\mathcal{A}^a_{f,i}(\mathcal{U}_i)\bigg\}+\ldots \
,\label{fa}
\end{eqnarray}
where the dots refer to possible additional corrections from other
sectors. Here, $E(U,k)$ are the non-holomorphic Eisenstein series
of order $k$, whereas $\mathcal{A}^a_{f,i}$ is defined as in
(\ref{index}), but replacing the regularized Eisenstein series
$\hat E_2$, appearing inside $D_{10}E_{10}$, by the holomorphic
one, $E_2$. The argument is defined as
$\mathcal{U}_i=(j+ipU_i)/k$, corresponding to the complex
structure induced in the worldvolume of the E1-instantons
\cite{kirit,kirit2}, as will be clear below.

Several comments for the $T^4/\mathbb{Z}_2$ orbifold limit of K3,
where the type I/SO(32) heterotic duality is explicit, are in
order:
\begin{itemize}
\item \emph{One-loop $\alpha'$ corrections to the K\"ahler
potential and the gauge kinetic function.} These are given by
eq.(\ref{kloop}) and the logarithmic term in eq.(\ref{fa}). These
expressions were also obtained by direct computation in the type I
side in \cite{fabre,haack}. In our context, these corrections come
respectively from non-holomorphic and holomorphic terms in
contributions of degenerate orbits under the modular group.

\item \emph{Tree-level contribution to the gauge kinetic
function.} Because D9-branes are fractional, the gauge kinetic
function receives an extra contribution proportional to,
\begin{equation}
\sim \ \sum \sqrt{\textrm{det}(P[G+F_2])}\ T\ ,
\end{equation}
where the sum runs over the $\mathbb{C}^2/\mathbb{Z}_2$
singularities and $P[\ldots]$ is the pull-back to the collapsed
2-cycle of the singularity. In the orbifold limit, the volume of
the 2-cycle is zero and the contribution from the metric vanishes.
However, the U(1) instantons hidden at the singularities
\cite{polcho} lead to a non-trivial linear term in $T$ in the
gauge kinetic function of the D9-branes.

\item \emph{Non-perturbative $E1$ multi-instanton corrections to
the K\"ahler potential and the gauge kinetic function.} They
appear as the Polyakov action of an instanton wrapping $N=kp$
times the $i$-th 2-torus, weighted by a modular function of
$\mathcal{U}_i$ and summed over $N$ and all the possible
wrappings.

In order for the instantons to contribute to the gauge kinetic
function, only four fermionic neutral zero modes should be
massless (corresponding to the ``goldstinos'') \cite{blumen}. A
$U(1)$ instanton on top of a singularity behaves as a ``gauge''
instanton for the U(1) gauge theory inside the corresponding half
D5-brane. These instantons are analogous to the ones discussed in
\cite{petersson}, with the extra fermionic zero modes lifted by
couplings with the D5-branes (see also
 \cite{billo2}).
Therefore they should be responsible for the 1-instanton
contribution. Notice however that in this case there is also a
Higgs branch which consists of moving the instanton out of the
singularity, leading to a SO(1) instanton (plus its image under
the orbifold). In this limit, the instanton has too many zero
modes and does not correct the gauge kinetic function.

Similarly, for the $N$-instanton contribution, in a generic point
of the moduli space the instanton gauge group is $SO(1)^{N}$, and
has too many zero modes. Only in the special locus on which all
the components of the multi-instanton are on top of the
singularity and the gauge group is enhanced to $U(N)$, four zero
modes survive, with the extra zero modes presumably lifted by
interactions with the D5-branes. Explicit examples where
instantons only contribute in a given locus of their moduli space
have been discussed in detail in the recent literature (see e.g.
\cite{cvetic,angel,angel2}).
\end{itemize}

\section{OUTLOOK}

Understanding the leading corrections to the low energy effective
theory in generic compactifications would constitute an important
input for addressing phenomenological and cosmological questions
within String Theory. It would be therefore very interesting to
extend these results to more involved and phenomenologically
relevant String Theory setups. For that, a better understanding of
the blow-up modes or the NS5-brane in heterotic string theory are
certainly desirable. We hope that the effort along those lines
will continue to be pursued in the near future.

\begin{acknowledgments}
I thank my collaborator E.~Dudas, for illuminating discussions and
ideas during the elaboration of \cite{cd} and afterwards. Also I
thank M. Lennek and M. Trapletti for useful comments, and the
organizers of the ICHEP'08 conference, in particular those of the
Formal Theory session, where this talk was presented. This work is
supported by an Individual Marie-Curie IEF. Additional support
comes from the contracts ANR-05-BLAN-0079-02, MRTN-CT-2004-005104,
MRTN-CT-2004-503369, MEXT-CT-2003-509661 and CNRS PICS \#~2530,
3059, 3747.
\end{acknowledgments}

\end{document}